\documentclass[11pt]{article}
\pdfoutput=1
\usepackage{amsmath,amssymb,color,epsfig,cite}
\usepackage{graphicx}
\usepackage{subfigure}
\usepackage{setspace}
\usepackage{amsthm,mathenv}

\allowdisplaybreaks[4]

\def\baselinestretch{1.4}
\usepackage{amsfonts}

\makeatletter
\@addtoreset{equation}{section}
\makeatother

\newcommand{\be}{\begin{equation}}
\newcommand{\ee}{\end{equation}}
\newcommand{\bea}{\setlength\arraycolsep{2pt} \begin{eqnarray}}
\newcommand{\eea}{\end{eqnarray}}
\newcommand{\nn}{\nonumber}

\def\0{{\sst{(0)}}}
\def\1{{\sst{(1)}}}
\def\2{{\sst{(2)}}}
\def\3{{\sst{(3)}}}
\def\4{{\sst{(4)}}}
\def\5{{\sst{(5)}}}
\def\6{{\sst{(6)}}}
\def\7{{\sst{(7)}}}
\def\8{{\sst{(8)}}}
\def\sst#1{{\scriptscriptstyle #1}}

\begin{document}

\begin{center}
{\large {\bf Stability of the Einstein Static Universe in $4 D$ Gauss-Bonnet Gravity}}

\vspace{15pt}
{\large Shou-Long Li,  Puxun Wu and Hongwei Yu}

\vspace{15pt}

{\it Department of Physics and Synergetic Innovation Center for Quantum Effect and Applications, Hunan Normal University, Changsha 410081, China}

\vspace{40pt}

\underline{ABSTRACT}
\end{center}
By rescaling the Gauss-Bonnet (GB) coupling constant $\alpha \rightarrow \alpha/(D-4)$ and considering the $D \rightarrow 4$ limit, the GB gravity gives rise to nontrivial modification of general relativity in four dimensions. In this work, we investigate the realization of the emergent universe scenario in the $4 D$ GB gravity. First, we obtain the Einstein static universe filled with a perfect fluid. Then, we show that  both spatially closed and open universes can be stable against both homogeneous and inhomogeneous scalar perturbations simultaneously.

\vfill
 shoulongli@hunnu.edu.cn \ \ \ pxwu@hunnu.edu.cn \ \ \ hwyu@hunnu.edu.cn


\thispagestyle{empty}

\pagebreak



\newpage

\section{Introduction}\label{sec1}
According to the standard big bang cosmology, our Universe began with a big bang and evolved from a nearly infinitely hot dense initial state. By postulating that the universe expands exponentially in the very early stage of its evolution, inflationary cosmology~\cite{Smeenk:2018dbt, Brout:1977ix, Starobinsky:1980te, Sato:1980yn, Fang:1980wi}  explains the horizon problem, flatness problem and magnetic-monopole problem very well. However, the singularity problem still remains one of the most prominent  open problems in the standard model of cosmology~\cite{Hawking:1969sw}. 
In order to avoid the initial singularity, many potential solutions have been proposed to provide a nonsingular cosmological model. One class of the solutions is a scenario which assumes that  the universe undergoes a transition from a contracting phase to an expanding phase~\cite{Finelli:2001sr, Gasperini:1992em, Khoury:2001wf, Brandenberger:1988aj, Steinhardt:2001st, Feng:2018qnx}. Another class  is the past-eternal universe which is a nonsingular cosmological model without a bounce~\cite{Bondi:1948qk, Hoyle:1948zz, Ellis:2002we, Ellis:2003qz}. One important scenario of the past-eternal universe is the so-called emergent universe  proposed by Ellis et al.~\cite{Ellis:2002we, Ellis:2003qz}, which assumes that  initially,  our Universe starts from an Einstein static universe with finite size, and then evolves to an inflationary era. This scenario has many attractive advantages. For instance, the horizon problem can be solved before inflation begins. In addition, there is no singularity, no exotic physics is involved, and the quantum gravity regime can even be avoided. Furthermore, it has been argued that the Einstein static state is favored by entropy considerations as the initial state for our universe~\cite{Gibbons:1987jt, Gibbons:1988bm, Mulryne:2005ef}.

The Einstein static universe was first proposed  by Einstein to describe a homogeneous and isotropic closed universe which neither expands nor contracts. However, it was soon demonstrated by Eddington~\cite{Eddington:1930zz} that the solution is unstable with respect to homogeneous and isotropic scalar perturbations in general relativity (GR).  On the other hand, it was also found that the universe is expanding rather than static. So, the mainstream cosmological model for the evolution of our universe  was replaced by a dynamical model described Friedmann-Robertson-Walker~(FRW) metric. Recently, The Einstein static universe was revived in the emergent universe scenario to avoid the big bang singularity. It should be pointed out that the  stability of the Einstein static universe is crucial to a successful realization of the scenario. Actually, it was found that the Einstein static universe is stable against small inhomogeneous vector and tensor perturbations as well as adiabatic scalar density perturbations if the universe contains a perfect fluid with $w = c_s^2 >1/5$ in GR~\cite{Gibbons:1987jt, Gibbons:1988bm, Barrow:2003ni}. The stability of the Einstein static universe has also been extensively studied in vast modified gravities, for example, GR with an extra field~\cite{delCampo:2007mp, delCampo:2009kp, Huang:2014fia,Huang:2018kqr, Miao:2016obc, Atazadeh:2015zma, Parisi:2012cg, Li:2019laq, Mousavi:2016eof}, higher derivative gravity~\cite{Boehmer:2007tr, Seahra:2009ft, Goheer:2008tn, Wu:2011xa, Li:2013xea, Paul:2008id, Bohmer:2009fc, Huang:2015kca, Wu:2009ah, Boehmer:2009yz, Heydarzade:2015hra, Goswami:2008fs}, high dimensional gravity~\cite{Zhang:2010qwa, Gergely:2001tn, Atazadeh:2014xsa, Zhang:2016obw} and so on~\cite{Boehmer:2013oxa, Huang:2015zma, Boehmer:2003iv, Canonico:2010fd, Carneiro:2009et,Atazadeh:2016yeh, Li:2017ttl, Parisi:2007kv, Yu:2019cku,Zhang:2013ykz}. 

Among these modified gravities, the Gauss-Bonnet (GB) gravity is the simplest modification of general relativity to include higher derivative curvature terms, while the resulting equation of motion (EOM) is still second order and has been proven to be free of ghosts~\cite{Lovelock:1971yv}. In past years, the GB gravity has been extensively studied in many aspects, such as  black holes and holography, e.g.~\cite{Boulware:1985wk, Cai:2001dz, Cvetic:2001bk, Brigante:2007nu, Pan:2009xa}. Since the GB combination is a topological invariant and plays no role in four dimensions,  the original GB gravity has been rarely studied in cosmology directly. Generally, the GB gravity was considered in high dimensional brane cosmology~\cite{Charmousis:2002rc}, and modified by the function $f({\cal G})$~\cite{Nojiri:2005jg}  or coupled with scalar fields~\cite{Nojiri:2005vv}. The emergent universe scenario has also been studied in the modified GB gravity~\cite{ Bohmer:2009fc, Huang:2015kca} and the GB gravity coupled with dilatons~\cite{Paul:2008id}. However, the Einstein static universe is unstable when both homogeneous and inhomogeneous scalar perturbations are taken into account although it is stable against only homogeneous or inhomogeneous perturbations. 

Recently, the GB gravity was revived in four dimensions by rescaling the GB coupling constant and considering the $D\rightarrow 4$ limit~\cite{Glavan:2019inb}. The similar idea was also considered  before motivated by the study of quantum corrections~\cite{Tomozawa:2011gp, Cognola:2013fva}. Many new predictions have been studied in the literature~\cite{Glavan:2019inb, Konoplya:2020bxa, Guo:2020zmf,  Fernandes:2020rpa, Casalino:2020kbt, Wei:2020ght, Konoplya:2020qqh, Kumar:2020owy, Hegde:2020xlv, Doneva:2020ped, Ghosh:2020vpc, Zhang:2020qew, Lu:2020iav, Singh:2020xju, Konoplya:2020ibi, Ghosh:2020syx, Konoplya:2020juj,  Kobayashi:2020wqy, Zhang:2020qam, HosseiniMansoori:2020yfj, Kumar:2020uyz, Singh:2020nwo, Wei:2020poh, Churilova:2020aca, Kumar:2020xvu, Islam:2020xmy,Liu:2020vkh, Heydari-Fard:2020sib, Konoplya:2020cbv, Jin:2020emq, Ai:2020peo, Zhang:2020sjh,EslamPanah:2020hoj, NaveenaKumara:2020rmi}. While the exact theory of the so called $4 D$ GB theory remains controversial~\cite{Gurses:2020ofy}, one concrete approach is to apply the Kaluza-Klein procedure, keeping consistently only the four-dimensional metric, together with the breathing mode.  One can then consistently set the dimensions $(D-4)$ of the internal space to zero,  leading to a scalar-tensor theory of the Hordenski class~\cite{Lu:2020iav}. This means that the procedure of the $D\rightarrow 4$ limit is not a simple trick.
So, a question arises naturally as to whether the emergent universe scenario can be realized in the  GB gravity by rescaling the GB coupling constant and considering the $D\rightarrow 4$ limit, against both homogeneous and inhomogeneous scalar perturbations simultaneously. We will answer this question in this paper.

 The remaining part of this paper is organized as follows. In Sec.~\ref{sec2}, we obtain the Einstein static universe  in the context of  the $4 D$ GB gravity with a perfect fluid. In Sec.~\ref{sec3}, we study the stability against both homogeneous and inhomogeneous scalar perturbations for spatially closed and open universes respectively.  Finally, we conclude  in Sec.~\ref{sec4}.


\section{Einstein static universe in $4D$ GB gravity}\label{sec2}

In this section, we will firstly briefly review the Friedmann equations in the GB gravity by rescaling the GB coupling constant and considering the $D\rightarrow 4$ limit. Then, we will give the Einstein static universe solution.
The action ${\cal S}$ of the $D$-dimensional GB gravity is given by
\be
{\cal S} =  \int d^D x \sqrt{-g} ( R + \alpha {\cal G } )  +{\cal S}_m \,,
\ee
where 
\be
{\cal G}= R^2 -4 R_{\mu\nu}R^{\mu\nu} +R_{\mu\nu\rho\sigma} R^{\mu\nu\rho\sigma} \,,
\ee
is the GB term, $\alpha$ is the coefficient of coupling constant and is positive in the heterotic string theory,  and $g$ and ${\cal S}_m$ respectively represent  the determinant of $g_{\mu\nu}$ and the action of matters. By varying $g^{\mu\nu}$, the EOM is given by
\be
G_{\mu\nu} + \alpha (-\frac12 g_{\mu\nu} {\cal G} +2 R_{\mu\lambda\rho\sigma} {R_\nu}^{\lambda\rho\sigma} -4 R_{\mu\lambda} {R_\nu}^\lambda -4 R_{\mu\rho\nu\sigma} R^{\rho\sigma} +2 R R_{\mu\nu}) =  T_{\mu\nu} \,, \label{eom1}
\ee
where $G_{\mu\nu}$ is the Einstein tensor, and  $T_{\mu\nu} = -\frac{1}{\sqrt{-g}}\frac{\partial {\cal S}_m}{\partial g^{\mu\nu}}$ is the energy-momentum tensor. The trace of the above equation is
\be
\frac{D-2}{2} R + \frac{(D-4) \alpha}{2} {\cal G} = {T^\mu}_\mu \,. \label{eom2}
\ee
First, we consider the $D$-dimensional FRW metric
\be
ds^2 = -dt^2 +a(t)^2 d\Omega_{D-1, \kappa}^2 \,, \label{nfrw}
\ee
where $a $ is the scale factor, $d\Omega_{D-1, \kappa}^2$ is the line element of $(D-1)$-dimensional maximally symmetric space  which can be written as
\be
d\Omega_{D-1, \kappa}^2 =\eta_{ij} dx^i dx^j =  \frac{dr^2}{1-\kappa r^2} +r^2 d\sigma^2 \,,
\ee
and $\kappa = 1, 0, -1$ denote the metric for the closed, flat and open universe respectively.  We assume that the matter is a perfect fluid, and the corresponding  energy-momentum tensor is given by
\be
T^{\mu\nu} = (\rho + P) u^\mu u^\nu +P g^{\mu\nu} \,, \quad \textup{with} \quad P = w \rho \,, \label{fluid}
\ee
where $\rho$ and $P$ represent energy density and pressure respectively, $w$ is the constant equation-of-state parameter, and velocity $u^\mu$ reads
\be
u^\mu = \left(1, 0, \dots, 0\right) \,, \quad \textup{satisfying} \quad u^\mu u_\mu = -1 \,. \label{velocity}
\ee
Substituting Eqs.(\ref{nfrw}) and (\ref{fluid}) into EOMs (\ref{eom1}) and (\ref{eom2}), we have the $D$-dimensional Friedmann equations 
 \bea
&& -\frac{(D-1)(D-2) (\kappa +\dot{a}^2 )}{2 a^2} - \frac{(D-1)(D-2)(D-3)(D-4)\alpha (\kappa +\dot{a}^2)^2}{2 a^4} = -\rho \,,\\
&&  -\frac{(D-1)(D-2)(D-3)(D-4)\alpha ((D-4)(\kappa +\dot{a}^2) +4 a \ddot{a})}{2 a^4} \nn \\
&&-\frac{(D-1)(D-2)((D-2)(\kappa+\dot{a}^2 ) +2 a\ddot{a} )}{2 a^2} = -\rho +(D-1) P \,.
 \eea
It is easy to see that the Gauss-Bonnet term vanishes in four dimensions. Now following  Ref.~\cite{Glavan:2019inb}, we consider 
\be
\alpha = \frac{\tilde{\alpha}}{D-4} \,, \label{rescale}
\ee
and then we adopt $D=4$. The metric reduces to the standard FRW metric, and the corresponding four-dimensional Friedmann equations become
\bea
&& \frac{3 \left(\dot{a}^2+\kappa \right)}{a^2} +\frac{3 \tilde{\alpha}  \left(\dot{a}^2+\kappa \right)^2}{a^4} =\rho \,,\\
&& -\frac{3 \left(2 a \ddot{a}+2 \left(\dot{a}^2+\kappa \right)\right)}{a^2} -\frac{12 \tilde{\alpha}  \ddot{a} \left(\dot{a}^2+\kappa \right)}{a^3}= -\rho +3 P \,,
\eea
where the dot represents the derivative with respect to time.
For the purpose of obtaining the Einstein static universe,  we let the scale factor $a(t)=a_0 = \textup{const}.$, and then the FRW metric reduces to the metric of the Einstein static universe
\be 
ds^2 = -dt^2 +a_0^2 \left[\frac{dr^2}{1-\kappa r^2} +r^2 \left(d\theta^2 +\sin^2\theta d\phi^2\right)\right] \,. \label{esu}
\ee
The solution is given by
\be
a_0 = \sqrt{\frac{{\tilde \alpha } \kappa (1-3 w)}{3 w+1}} \,,\qquad \rho = \frac{6 (3 w+1)}{{\tilde \alpha } (1-3 w)^2} \,.
\ee
First, it is easy to see that the solution does not exist in pure GR. Second, there is no Einstein static flat universe ($\kappa = 0$) in the four-dimensional GB theory filled with a perfect fluid. So we only discuss the  closed ($\kappa =1$) and open ($\kappa = -1$) Einstein static universes.


\section{Stability analysis}\label{sec3}


Now we study the stability of the Einstein static universe in the $4 D$ GB gravity. First we will study the linear EOMs in $D$ dimensions. Then we perform the $D\rightarrow 4$ limit. We use the symbols bar and tilde to represent background and perturbation components respectively.  The perturbed metric can be written as
\be
g_{\mu\nu} = \bar{g}_{\mu\nu} + h_{\mu\nu} \,,
\ee
where $\bar{g}_{\mu\nu}$ is the background metric which is given by~Eq.~(\ref{nfrw}) with $a = a_0$ and the $h_{\mu\nu}$ is a small perturbation.
For our purpose, we consider scalar perturbations in the Newtonian gauge in which ${h_\mu}^\nu$ is given by
\be
{h_\mu}^\nu = \textup{diag} \left(-2 \Psi \,, 2 \Phi \,, \dots \,, 2 \Phi\right) \,. \label{pert}
\ee
Now the indexes are lowered and raised by the background metric. By using the relation $g^{\mu\nu} g_{\nu\lambda} ={\delta^\mu}_\lambda$, the inverse metric is perturbed by
\be
\widetilde{g}^{\mu\nu} = -\bar{g}^{\mu\rho} \bar{g}^{\nu\sigma} h_{\rho\sigma} \,.
\ee
For a perfect fluid, the perturbations of energy density and pressure are $\widetilde{\rho}$ and $\widetilde{P} = w \widetilde{\rho}$ . The perturbations of velocity are given by
\be
  \widetilde{u}_0 = \widetilde{u}^0 = \frac{h_{00}}{2}   \,, \quad \widetilde{u}^i = \bar{g}^{ij} \, \widetilde{u}_j = \bar{g}^{ij} \bar{\nabla}_j U  \,,
\ee
where ``\,0\," and ``\,$i, j$\," denote time and space components respectively. The perturbed energy momentum tensor is
\be
\widetilde{T}^{\mu\nu} = P \, \widetilde{g}^{\mu\nu} +\widetilde{P} \, \bar{g}^{\mu\nu} +  (\widetilde{\rho} + \widetilde{P}) u^\mu u^\nu +(\rho + P) \widetilde{u}^\mu u^\nu +(\rho + P) u^\mu \widetilde{u}^\nu  \,,
\ee
where $u^\mu , \rho , and P$ represent the unperturbed components. Considering above expressions, the linearized equations of Eqs.~(\ref{eom1}) and (\ref{eom2}) become
\bea
&&\widetilde{G}_{\mu\nu} + \alpha (-\frac12 h_{\mu\nu} {\cal G} -\frac12 g_{\mu\nu} {\cal \widetilde{G}} +2 \widetilde{R}_{\mu\lambda\rho\sigma} {R_\nu}^{\lambda\rho\sigma} +2 R_{\mu\lambda\rho\sigma} {\widetilde{R}_\nu}^{\lambda\rho\sigma} -4 \widetilde{R}_{\mu\lambda} {R_\nu}^\lambda -4 R_{\mu\lambda} {\widetilde{R}_\nu}^\lambda \nn \\
&& -4 \widetilde{R}_{\mu\rho\nu\sigma} R^{\rho\sigma}  -4 R_{\mu\rho\nu\sigma} \widetilde{R}^{\rho\sigma} +2 \widetilde{R} R_{\mu\nu}+2 R \widetilde{R}_{\mu\nu}) =  \widetilde{T}_{\mu\nu} \,, \label{lineom1}\\
&& \frac{D-2}{2} \widetilde{R} + \frac{(D-4) \alpha}{2} {\cal \widetilde{G}} = {\widetilde{T}^\mu}_\mu \,.  \label{lineom2}
\eea
$\Psi, \Phi, \widetilde{\rho}$ and $U$ are functions of $(t, r, x_1, \dots, x_{D-2})$. For scalar perturbations, it is useful to perform a harmonic decomposition~\cite{Harrison:1967zza},
\bea
\Psi = \Psi_n(t)\, Y_n (r, x_1,\dots, x_{D-2}) \,,\quad \Phi = \Phi_n(t) \,Y_n (r, x_1,\dots, x_{D-2}) \,,\nn \\
\widetilde{\rho} = \rho \, \xi_n(t) \,Y_n (r, x_1,\dots, x_{D-2}) \,,\quad U = U_n(t) \,Y_n (r, x_1,\dots, x_{D-2}) \,.\label{decomp}
\eea
In these expressions, summations over co-moving wavenumber $n$  are implied. The harmonic function $Y_n (r, x_1,\dots, x_{D-2})$ satisfies~\cite{Harrison:1967zza}
\bea
\Delta Y_n (r, x_1,\dots, x_{D-2}) &=& \frac{1}{\sqrt{-\det{\eta}}}\frac{\partial}{\partial \xi^i} \left(\sqrt{-\det{\eta}} \eta^{ij} \frac{\partial  }{\partial \xi^j} Y_n (r, x_1,\dots, x_{D-2})\right) \nn \\
&=&  -k^2 Y_n (r, x_1,\dots, x_{D-2}) \,,
\eea
where $\Delta$ is the Laplacian operator, $\xi^i = (r, x_1,\dots, x_{D-2})$, and $k$ is a separation constant. For a spatially closed universe corresponding to $\kappa =1$, we have $k^2 =n(n+2)$ where the modes are discrete $(n= 0, 1, 2\dots)$. For a spatially open universe corresponding to $\kappa =-1$, we have  $k^2 =n^2+1$ where $n\ge 0$. Formally, $n = 0$ gives a spatially homogeneous mode and $n = 1, 2\cdots$ correspond to spatially inhomogeneous modes for both closed and open universes~\cite{Barrow:2003ni, Huang:2015kca}.
Substituting Eqs.~(\ref{pert}) and (\ref{decomp}) into (\ref{lineom1})-(\ref{lineom2}), after some algebra, we have the $D$-dimensional effective perturbed equations,
\bea
&&\frac{2 {\alpha}  (D-4) (D-3) (D-2) \kappa  \left(-a_0^2 (D-1) \ddot{\Phi}_n -(D-4) \Phi_n  \left(-D \kappa +\kappa +k^2\right)+k^2 \Psi_n \right)}{a_0^4} + \nn \\
&& \frac{(D-2) \left(k^2 \Psi_n -a_0^2 (D-1) \ddot{\Phi}_n -(D-2) \Phi_n  \left((1-D) \kappa +k^2\right) \right)}{a_0^2} =\rho  \xi_n ((D-1)  w -1) \,, \\ 
&&-\frac{\alpha  (D-4) (D-3) (D-2) \kappa  \left((D-1) \kappa  \Psi_n -2 \Phi_n  \left(-D \kappa +\kappa +k^2\right)\right)}{a_0^4} \nn\\ 
&&+\frac{(D-2) \left(\Phi_n \left(-D \kappa +\kappa +k^2\right)+(\kappa -D \kappa ) \Psi_n \right)}{a_0^2}+\rho  (2 \Psi_n -\xi_n )=0 \,,\\
&&a_0^2 \rho  (w+1) U_n -(D-2) \dot{\Phi}_n \left(a_0^2+2 \alpha  (D-4) (D-3) \kappa \right)  =0 \,, \\
&&\Psi_n  \left(a_0^2+2 \alpha  (D-4) (D-3) \kappa \right)-(D-3) \Phi_n  \left(a_0^2+2 \alpha  (D-4) (D-5) \kappa \right) =0 \,.
\eea
Now we rescale $\alpha$ as Eq.~(\ref{rescale}) and adopt $D= 4$, and then the corresponding perturbed equations are given by 
\bea
&&\frac{2 k^2 \Psi  \left(2 \tilde{\alpha}  \kappa +a_0^2\right)}{a_0^4}-\frac{4 \left(k^2-3 \kappa \right) \Phi }{a_0^2}+\left(-\frac{12 \alpha  \kappa }{a_0^2}-6\right) \ddot{\Phi} = \xi \rho  (3 w -1 ) \,, \\ 
&&-\frac{2 \tilde{\alpha}  \kappa  \left(3 \kappa  \Psi -2 \left(k^2-3 \kappa \right) \Phi \right)}{a_0^4}+\frac{2 \left(\left(k^2-3 \kappa \right) \Phi -3 \kappa  \Psi \right)}{a_0^2}+\rho  (2 \Psi -\xi ) =0 \,,\\
&&a_0^2 \rho  (w+1) U -2 \left(2 \tilde{\alpha } \kappa +a_0^2\right) \dot{\Phi } =0 \,, \\
&&\Psi \left(2 \tilde{\alpha}  \kappa +a_0^2\right)-\Phi  \left(a_0^2-2 \alpha  \kappa \right) =0 \,.
\eea
We can solve the equations to get
\bea
&&U_n =\frac{2 \left(2 \tilde{ \alpha}  \kappa +a_0^2\right) \dot{\Phi}_n }{a_0^2 \rho  (w+1)} \,, \\
&&\xi_n = -\frac{2 \Phi_n  \left(a_0^2 \left(k^2-6 \kappa \right)+2 \tilde{ \alpha}  \kappa  k^2\right)+6 a_0^2 \left(2 \tilde{ \alpha}  \kappa +a_0^2\right) \ddot{\Phi }_n }{a_0^4 \rho  (3 w-1)} \,, \\
&&\Psi_n = \frac{\Phi_n  \left(a_0^2-2 \tilde{ \alpha}  \kappa \right)}{2 \tilde{ \alpha}  \kappa +a_0^2} \,,
\eea
where $\Phi_n$  satisfies a second order ordinary differential equation
\be
\ddot{\Phi}_n +  Z \Phi_n =0 \,, \label{ode}
\ee
in which the dot represents the derivative with respect to time and 
\be
Z = -\frac{(3 w+1) \left(\kappa +3 k^2 w (w+1)-9 \kappa  w^2\right)}{3 \tilde{ \alpha}  \kappa  (w+1) (3 w-1)} \,.
\ee
To analyze the stability of the Einstein static universe in the GB gravity, we need to discuss the condition of existence of the oscillating solution of Eq.~(\ref{ode}). The condition is
\be
Z>0  \,.
\ee
Apart from this condition, we should require the coupling constant $\tilde{\alpha}$, the energy density of the perfect fluid $\rho$ and $a_0$ satisfy
\be
\tilde{\alpha} >0 \,, \quad \rho>0   \quad \textup{and} \quad a_0 > 0\,. 
\ee

In the case of $\kappa =1$, which corresponds to the Einstein static closed universe, the stability requires 
\bea
&& k^2 = 0 \,, \quad -\frac13 \le w \le \frac13 \,,\\
&& k^2\ge 8 \,, \quad f(k^2)<w<\frac{1}{3} \,,\quad \textup{with} \quad f(k^2) = \frac{\sqrt{9 k^4-12 k^2+36}-3 k^2}{6 \left(k^2-3\right)} \,.
\eea
Note that $f(k^2)$ is a monotonic increasing function, and as  $k^2 \rightarrow \infty$,  $f(k^2) \rightarrow 0$. As a result,  we have 
\be
-\frac13 < f(k^2\ge 8) < 0  \,.
\ee
So, the Einstein static closed universe can be stable against both the homogeneous ($k^2 = 0$) and inhomogeneous ($k^2 \ge 8$) scalar perturbations simultaneously under the condition
\be
 0 \le w<\frac{1}{3} \,.
 \ee
We find that the stable Einstein static closed universe filled with non-relativistic matter and  radiation does exist.

In the case of $\kappa = -1$, which corresponds to the Einstein static open universe, the stability requires 
\be
 w>\frac13 \,.
\ee

\section{Conclusion and discussion}\label{sec4}

We investigate the realization of the emergent universe scenario in the GB gravity by rescaling the GB coupling constant and considering the $D \rightarrow 4$ limit. 
First, we obtain the Einstein static universe filled with a perfect fluid in the $4 D$ GB gravity. Then, we find the stability conditions against both homogeneous and inhomogeneous scalar perturbations simultaneously  for spatially closed and open universes respectively. We find that the Einstein static universe with  closed spatial geometry can be stable against both the homogeneous and inhomogeneous scalar perturbations simultaneously. Such a stable universe is filled with ordinary matter and radiation. The Einstein static universe with  open spatial geometry can be also stable.

Finally, we want to point out that our work only shows that  the emergent universe scenario may be realized in the GB gravity by rescaling the GB coupling constant and considering the $D \rightarrow 4$ limit since the Einstein static universe is stable against scalar perturbations. The stabilities against vector and tensor perturbations are also worthy of investigation. Besides, it is also worth studying the stabilities in the concrete scalar tensor theory given in Ref.~\cite{Lu:2020iav}. Finally, there are also many other important issues in cosmology which should be studied in the novel four-dimensional theory. We leave these to the future work.


\section*{ACKNOWLEDGEMENTS}
We are grateful to Profs.~H. L\"u and Hao Wei for kind and useful discussions.  This work was supported in part by  NSFC grants No. 11435006, No. 11690034, No. 11775077, No. 11947216 and China Postdoctoral Science Foundation 2019M662785.

\renewcommand{\baselinestretch}{1.0}


\end{document}